\documentclass[sigconf,screen, nonacm]{acmart}

\renewcommand\footnotetextcopyrightpermission[1]{} 
\usepackage{xcolor}
\usepackage{ragged2e}
\usepackage{natbib}
\usepackage{amsmath,amsfonts}
\usepackage{algorithmic}
\usepackage{graphicx}
\usepackage{textcomp}
\def\BibTeX{{\rm B\kern-.05em{\sc i\kern-.025em b}\kern-.08em
    T\kern-.1667em\lower.7ex\hbox{E}\kern-.125emX}}

\usepackage[]{hyperref}
\usepackage{tcolorbox}
\usepackage{listings}
\usepackage{letltxmacro}

\usepackage[font=normalfont]{caption}
\usepackage{setspace}
\usepackage[labelfont=normalfont]{subcaption}

\LetLtxMacro\oldttfamily\ttfamily
\DeclareRobustCommand{\ttfamily}{\oldttfamily\csname ttsize\endcsname}

\definecolor{codegray}{rgb}{0.5,0.5,0.5}
\definecolor{codepurple}{rgb}{0.58,0,0.82}
\definecolor{backcolour}{rgb}{0.95,0.95,0.92}
\definecolor{codegreen}{rgb}{0,0.6,0}
\definecolor{codebg}{RGB}{230,230,230}
\usepackage{tikz}
\usepackage{amsmath}

\usepackage{multirow}
\usepackage{rotating}

\AtBeginDocument{%
  \providecommand\BibTeX{{%
    Bib\TeX}}}

\setcopyright{acmlicensed}
\copyrightyear{2024}
\acmYear{2024}
\acmDOI{XXXXXXX.XXXXXXX}

\acmISBN{978-1-4503-XXXX-X/18/06}

\makeatletter
\def\@copyrightspace{\relax}
\makeatother

\setcopyright{none}

\begin{document}

\title{Shield Bash: Abusing Defensive Coherence State
Retrieval to Break Timing Obfuscation}  

\author{Kartik Ramkrishnan}
\affiliation{%
  \institution{University of Minnesota, Twin Cities}
  \city{Minneapolis}
  \state{Minnesota}
  \country{USA}
}
\author{Antonia Zhai}
\affiliation{%
  \institution{University of Minnesota, Twin Cities}
  \city{Minneapolis}
  \state{Minnesota}
  \country{USA}}
%
\author{Stephen McCamant}
\affiliation{%
  \institution{University of Minnesota, Twin Cities}
  \city{Minneapolis}
  \country{USA}
}
\author{Pen Chung Yew}
\affiliation{%
 \institution{University of Minnesota, Twin Cities}
 \city{Minneapolis}
 \state{Minnesota}
 \country{USA}}
%
%
%
%
%


\begin{abstract}
Microarchitectural attacks are a significant concern, leading to many hardware-based defense proposals. However, different defenses target different classes of attacks, and their impact on each other has not been fully considered. \textcolor{black}{To raise awareness of this problem}, we study an interaction between two state-of-the art defenses in this paper,  \emph{timing obfuscations of remote cache lines} (TORC) and \emph{delaying speculative changes to remote cache lines} (DSRC). TORC mitigates cache-hit based attacks and DSRC mitigates speculative coherence state change attacks.

We observe that DSRC enables coherence information to be retrieved into the processor core, where it is out of the reach of timing obfuscations to protect. This creates an unforeseen consequence that redo operations can be triggered within the core to detect the presence or absence of remote cache lines, which constitutes a security vulnerability. We demonstrate that a new covert channel attack is possible using this vulnerability. 
We propose two ways to mitigate the attack, whose performance varies depending on an application's cache usage. 
One way is to never send remote \emph{exclusive} coherence state (E) information to the core even if it is created.
The other way is to never create a remote E state, which is responsible for triggering redos.  

We demonstrate the timing difference caused by this microarchitectural defense assumption violation using GEM5 simulations. Performance evaluation on SPECrate 2017 and PARSEC benchmarks of the two fixes show less than 32\% average overhead across both sets of benchmarks. The repair which prevented the creation of remote E state had less than 2.8\% average overhead. 
\end{abstract}

\ccsdesc[500]{Security and privacy~Side-channel analysis and countermeasures}
\ccsdesc[500]{Computer systems organization~Architectures}

\keywords{Cache, security, side-channel, attack, defense, assumption, violation}


\maketitle

\section{Introduction}
\label{section:introduction}
With the increasing instances of microarchitectural attacks~\cite{gruss2016flush+,purnalprime+,gruss2015cache,kocher2019spectre,kocher2020spectre,lipp2018meltdown}, computer architects face an important challenge in designing attack-resilient architectures.
Many classes of microarchitectural attacks are launched through the last-level cache (LLC) shared among security domains running on different cores.
For example, an attacker process can spy on a victim process's last-level cache (LLC) accesses despite not having access to the private caches of the victim. Microarchitectural leakage can occur through speculative~\cite{kocher2019spectre} or non-speculative~\cite{gruss2016flush+} LLC accesses.

Diverse defenses~\cite{ramkrishnan2020first,ojha2021timecache,yan2019secdir,ainsworth2020muontrap,ainsworth2021ghostminion,yan2018invisispec,aimoniotis2023recon} have been proposed in response to these attacks.
\begin{figure}
  \includegraphics[width=\linewidth]{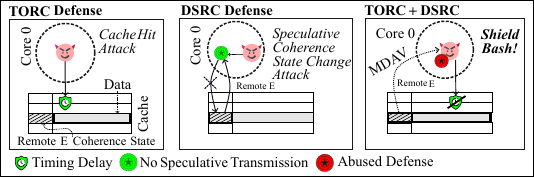}
  \\[-2ex]
  \caption{\textmd{TORC, DSRC and TORC+DSRC defenses, where an attacker in a remote Core 0 (dashed outline circle) accesses remote cache line data (shaded gray) with E coherence state.
  TORC mitigates a cache hit attack (green shield with timer) on victim Core 1. DSRC checks for remote E state and disallows speculative changes to it (green round shield). In TORC + DSRC, the attacker unexpectedly gets secret-related coherence feedback (an instance of an MDAV), enabling new shield-bash attacks and breaking timing obfuscation.}} 
  \label{fig:mdav_introduction}
\end{figure}
These defenses do not consider a broader  threat model that includes multiple kinds of attacks. 
Consequently, these defense mechanisms fail to consider whether their deployment would impact other defenses that are aimed to defend the system against  different attacks, leading to unforeseen circumstances. 
 For example, in many microarchitectures ~\cite{lowe2020gem5,ramkrishnan2020first,ojha2021timecache}, a cache hit access returns data to the core and not other information, e.g., coherence state. A defense  may understandably be designed while making an implicit assumption that this is the case. On the other hand, other defenses may make changes to the microarchitecture that render this assumption invalid.

In the presence of \emph{microarchitectural defense assumption violations} (MDAVs) we anticipate the possibility of scenarios where an attacker could `shield bash' through a victim's defenses by abusing defensive capabilities available in the hardware. 
This is analogous to a common media trope that a shield, usually used for defense against attacks, can also be used as a bludgeoning weapon~\cite{schaefer2013blizzard}. 

\textcolor{black} {As a concrete example of the above idea}, we consider two state-of-the-art defense strategies, namely, \emph{timing obfuscations of remote cache lines}  (TORC) and  \emph{delaying speculative changes to remote cache lines} (DSRC), used together with an underlying MESI coherence protocol.
TORC defends against attacks that try to detect cache hits on lines inserted by remote cores (we refer to these cache lines as remote cache lines). It delays the data return time from these cache lines so that it resembles a cache miss (see \S\ref{subsection:remote_and_local}). 
TORC 
protects both remote \emph{exclusive} (E) and remote \emph{shared} (S) state cache lines from cache hit-based detection~\cite{yan2019secdir,ramkrishnan2020first,ojha2021timecache} (see \S\ref{subsection:timing_obfuscation}).
DSRC triggers a \emph{delay and redo} operation on remote \emph{exclusive} (E) or \emph{modified} (M) state cache lines, so that speculative accesses to the remote cache lines can be delayed and redone later on (if on the correct path). DSRC is used as part of low-overhead speculative attack mitigation strategies~\cite{ainsworth2021ghostminion,ainsworth2020muontrap,saileshwar2019cleanupspec}
(see \S\ref{subsection:speculative_leakages}).  
It is a plausible choice to adopt both TORC and DSRC in a system for a defense that mitigates both attacks. However, DSRC returns coherence state to the core, which breaks a fundamental assumption of TORC that only data is returned to the core, not any additional metadata. 
Thus, it is possible for an attacker to break TORC's defense of cache state.
Figure~\ref{fig:mdav_introduction} shows how putting together TORC and DSRC creates an MDAV. 
 The left pane shows how TORC creates a timing delay (green shield) when an attacker (pink devil) accesses a remote E cache line, thus mitigating cache-hit attacks. 
 The middle pane shows how DSRC takes feedback from remote E coherence state (green round shield with a star) and prevents any speculative coherence state  changes.  
 The right pane shows that putting them together allows the attacker to retrieve the remote E state, leading to a violation in TORC's assumption about what information the attacker has access to.

We discuss the environment in which we mount this `shield bash' attack on TORC using DSRC in \S\ref{subsection:environment}. We illustrate how it is applied on a cache with a MESI-like coherence protocol using sharer bit-vectors, in \S\ref{sub:doing_attacks}.
\begin{figure}
\begin{subfigure}[b]{\linewidth}
  \includegraphics[width=\linewidth]{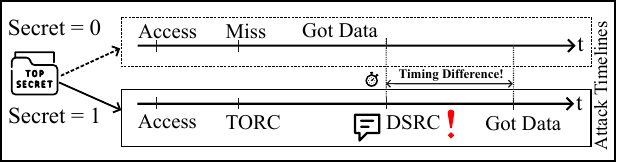}
  \\[-3ex]
  \caption{\textmd{The TORC + DSRC configuration shows a timing difference that can be observed by an attacker (shown as a stopwatch) based on the secret value being 0 or 1. This is due to a secret-dependent redo operation (shown by a red exclamation mark) triggered by DSRC's cache-core communication (shown as the chat icon).}}
  \label{fig:attack_overview}
\end{subfigure}
\\[1ex]

\begin{subfigure}[b]{\linewidth}
  \includegraphics[width=\linewidth]{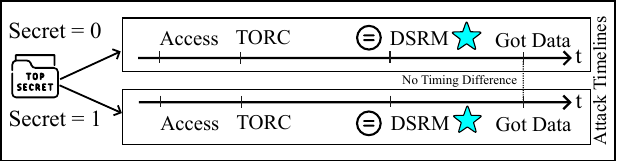}
  \caption{\textmd{The TORC + DSRM configuration shows no timing difference based on the secret value being 0 or 1. This is due to a secret-independent redo operation  (shown as a blue star) triggered by DSRM's equalized cache-core communication (shown as an equals-sign icon).}}
  \label{fig:defense_overview}
  \end{subfigure}
  \\[-2ex]
  \caption{\textmd{Timelines for the shield bash attack}}
  \end{figure}
\textcolor{black} {We propose two possible strategies, \emph{equalization} and \emph{elimination}, whose performance depends on the application's cache usage.} Our \emph{equalization} strategy sends the same coherence information to the core regardless of the actual presence or absence condition. The core conservatively triggers a redo in both cases for coherence state updates. We refer to this fix as \emph{delay speculative on remote and miss} (DSRM). 
\textcolor{black} {To clarify how the equalization works}, Figure~\ref{fig:defense_overview}  shows that the timing of TORC + DSRM has the same value regardless of the secret, thus mitigating the breach. This is due to a secret-independent redo operation (shown as a blue star) triggered by DSRM’s equalized cache-core communication (shown as an equalize icon), as opposed to a secret-dependent redo triggered by DSRC. DSRM and a simple optimization are both explained in \S\ref{subsection:mitigate}. 
\textcolor{black} {If the number of LLC misses is high, DSRM may have higher performance overheads. Another approach is to use a \emph{Start-With-S MESI} (or SS-MESI) strategy (instead of regular MESI) in conjuction with DSRC and TORC (eliminating unsafe E states). We always start with S state on load misses in the LLC, eliminating the possibility of redo attacks. 
The performance overhead depends on the number of stores rather than the LLC miss counts (see~\S\ref{subsection:non_dsrm})}. 
The implementation of TORC, DSRC, DSRM and SS-MESI is done on the last-level cache of a modern three-level cache hierarchy.

We implement our attack probe in C with inlined \texttt{x86\_64} assembly. The probe triggers a secret-dependent redo of a load  using a \emph{load in right-path branch shadow} (LRBS) approach. LRBS includes a branch instruction for creating speculation, a delay load before the branch and a redo triggering load after the branch (see \S\ref{section:attack}).  
Our attack probe runs on the GEM5 simulator with microprocesor configurations that include relevant combinations of TORC, DSRC, DSRM and Start-With-S MESI.  
DSRM and SS-MESI eliminate the timing difference created by the probe (see \S\ref{subsection:attack_experiments}). 

\textcolor{black}{We perform GEM5 simulations of the above configurations on SPECrate 2017 and PARSEC benchmarks to estimate the costs of the mitigations (see~\S\ref{section:performance}).}
 To summarize, we make the following contributions in the paper.
 
\begin{itemize}

\item
We introduce the concept of \emph{microarchitectural defense assumption violation} (MDAV). As an instance, we show that using  state-of-the-art defense strategies, DSRC and TORC, concurrently, causes an assumption violation by sending coherence information from the cache to the core. This can be abused to carry out a `shield bash' attack to break TORC using speculative loads as attack probes. We  propose and implement a \emph{load in right-path branch shadow} (LRBS) attack probe that uses the above idea to break TORC.

\item
\textcolor{black} {We propose two mitigation approaches targeting different aspects of the vulnerability. The first approach (equalization) prevents unsafe cache line presence information from reaching the core, which we refer to as a \emph{delay speculative on remote and miss} (DSRM) approach. The second approach (elimination) drops the E state in unsafe scenarios, eliminating the possibility of redos.}
\item
  \textcolor{black} {We implement DSRM and SS-MESI on top of TORC and DSRC defenses on a shared last-level cache.}  
We simulate the LRBS probe on a GEM5 simulator implementation and show that TORC is broken.

A GEM5 performance evaluation  using  SPECrate 2017 and PARSEC benchmarks shows that average performance overheads are less than 2.8\% for PARSEC workloads on average. The average performance overheads are less than 32\% (for DSRM) and 2.8\% (for T + DC + SS-MESI) for SPEC2017rate workloads. \textcolor{black} {In the majority of evaluated benchmarks, SS-MESI has a lower performance overhead, however, in a smaller number of cases, DSRM has lower overheads (see \S\ref{section:performance}).}  
\end{itemize}

 To the best of our knowledge, this paper is the \emph{first} work demonstrating an MDAV scenario due to remote E coherence information flowing from the cache to the core. \textcolor{black}{We also outline possible MDAVs between other defenses, such as speculative declassification, randomization and partitioning (see \S\ref{section:future_work}).}

 The rest of the paper is organized as follows.  \S\ref{section:background} contains the background for reading this paper. \S\ref{section:vulnerability} shows an example where the vulnerability is expoited to leak secret coherence state.
\S\ref{section:attack} presents a new attack probe against TORC. \S\ref{subsection:attack_experiments} presents simulation results for LRBS that show how TORC is broken. \S\ref{section:performance} presents the performance evaluation on SPECrate 2017 and PARSEC benchmarks.
\S\ref{section:discussion} discusses miscellaneous issues, related work and future work. \S\ref{section:conclusion} concludes the paper. Key acronyms are listed in Table~\ref{tab:acronyms}.

\section{Background}
\label{section:background}
First, we discuss the concept of remote and local cache lines in existing microarchitectures (see \S\ref{subsection:remote_and_local}). Second, we discuss two existing classes of microarchitectural attacks that include \emph{cache-hit based} microarchitectural attacks (see \S\ref{subsection:timing_obfuscation}) and \emph{speculative coherence state} microarchitectural attacks (see \S\ref{subsection:speculative_leakages}). We also discuss two state-of-the-art defense strategies for those two attacks, namely, TORC and DSRC respectively in the corresponding sub-sections (\S\ref{subsection:timing_obfuscation} and \S\ref{subsection:speculative_leakages}). Lastly, we discuss why TORC and DSRC need to be used together to mitigate both classes of attacks (see~\S\ref{subsection:need_for_both}).

\begin{table}[t]
\small
\resizebox{\linewidth}{!}{%
\begin{tabular}{ |p{1.3cm}|p{6.2cm} |
 }
 \hline
 \multicolumn{2}{|c|}{Acronym Expansion Reference} \\
 \hline
 \textbf{Acronym} & \textbf{Explanation} \\
 \hline
 MDAV & Microarchitectural defense assumption violation \\
 \hline
 TORC & Timing obfuscation of remote cache lines \\
 \hline
 DSRC & Delaying speculative changes to remote cache lines \\
 \hline
 DSRM & Delay speculative on remote and miss \\
 \hline
 LRBS & Load in right-path branch shadow \\
 \hline
\end{tabular}}
\\[2ex]
\caption{\textmd{Key acronyms defined in this work are summarized above. MDAV is the concept of defenses having assumption violations in the presence of other defenses. TORC and DSRC are existing defense. DSRM is a new updated version of DSRC. LRBS is an attack probe. }}
\label{tab:acronyms}
\end{table}

\subsection{Remote and Local Cache Lines}
\label{subsection:remote_and_local}
 
A \emph{remote} cache line is one that was not previously inserted into the cache hierarchy by the accessing core. 
A \emph{local} cache line is one that was inserted into the cache hierarchy by the accessor core at a previous time. Remote cache lines are a security concern because their presence or absence can reveal to an attacker a secret that was used by a victim  on a different core.

\begin{table}
\small
\begin{tabular}{ |p{2.4cm}| p{2.7cm} |  p{1cm} | p{0.9cm} |
 }
 \hline
 \multicolumn{2}{|c|}{\textbf{Testing Store Probes}}  & \multicolumn{2}{|c|}{\textbf{Cycles for Hit}}\\
 \hline
 \textbf{Architecture} & \textbf{Product Name}  & \textbf{S(L1)} & \textbf{E(L1)} \\
 \hline
 Haswell-E~\cite{haswell} & Xeon E5-2699 v3  & 137 & 65 \\
 \hline
 Icelake~\cite{icelake} &  Xeon Silver 4210  & 162  & 94 \\ 
 \hline

\end{tabular}
\\[2ex]
\caption{\textmd{We verify that store probes are a feasible strategy using experiments on two real machines with differing computer architectures. There is a significant timing difference between write hit of E/S states (cycles) for both machines. \textcolor{black} {This demonstrates the need for both TORC and DSRC defenses to mitigate leakages through these store-probes (see \S\ref{subsection:need_for_both}).}}} 
\label{tab:store_receiver_real_machine}
\end{table}

\subsection{Remote Presence/Absence Attacks and the TORC Defense}
\label{subsection:timing_obfuscation}

\textbf{Attack.}
 A classic example of a remote presence/absence  attack is against the modular exponentiation function used by a vulnerable implementation of RSA decryption~\cite{yarom2014flush+}. An attacker on a different core monitors remote cache line insertions into the cache to leak the secret key~\cite{yarom2014flush+, gruss2016flush+,gruss2015cache}. Another example is the monitoring of AES T-table accesses that leads to the leakage of secrets~\cite{seddigh2020enhanced}.
 Fast covert channels based on cache hit attacks are also a concern~\cite{saileshwar2021streamline}.

 The attacker typically has only read ability on the vulnerable cache line addresses. Therefore, the attacker detects the presence of those remote cache lines by timing the execution of load-based probes that access the vulnerable cache line address. If there is a cache hit (presence of a remote cache line), then the attack probe takes a shorter time to complete execution. If it is a cache miss (absence of a remote cache line), then it takes a longer time for the probe to complete execution.

\textbf{TORC Defense.}
Timing obfuscation is an important defense against remote presence/absence attacks~\cite{ramkrishnan2020first,ojha2021timecache,yan2019secdir}. The key idea is that any cache hits on remote cache lines are masked by increasing the total time taken for the data to return to the core. Thus, an attacker cannot differentiate the timing of a hit on a remote cache line (up-to-date data is found in the cache hierarchy) from that of a miss (up-to-date data is retrieved from the memory).

TORC defenses often leverage the sharer bit-vector~\cite{ramkrishnan2020first} or similar sharer information ~\cite{yan2019secdir} maintained by the coherence protocol in the LLC, to detect remote cache lines and to delay the access to make remote cache hits look like cache misses.

In most existing attack probes, there is only one cache/memory access for the load instruction accessing the vulnerable cache line address. Hence, TORC is effective because it mitigates this timing difference.
However, there can be multiple cache accesses generated by the load instruction  if there is a change in the underlying microarchitecture where TORC is deployed. 

\subsection{Speculative Coherence State Change Attacks and the DSRC Defense}
\label{subsection:speculative_leakages}
\textbf{Attack.} 
An example of an attack scenario where this technique is relevant involves an attacker manipulating the OS into first reading  a secret speculatively and then reading the user-space of an attacker process based on that speculatively read secret~\cite{ragab2024ghostrace}.
A possible cache side-effect that could be triggered using the above attack strategy is to  change the coherence state of a remote cache line that the attacker has placed into the cache previously (e.g., E/M$\to$S state transition of the MESI coherence protocol). The change in coherence state is a security issue because the access timings could change depending on the coherence state~\cite{yao2018coherence}.  When accessing a remote cache line, for example, $E$/$M$ state can take longer to access than $S$ state due to the timing overhead of accessing the remote private caches and forwarding the data from there, as opposed to direct retrieval of data for the $S$ state situation. The attacker detects the coherence state change using load-based probes typically~\cite{yao2018coherence}. We also note that store-based probes are effective, based on our tests on existing architectures. This is because the time needed to complete a store on an S state cache line (invalidate broadcast) is much larger than the time taken to complete a store on an E state cache line (silent upgrades).
Table~\ref{tab:store_receiver_real_machine} contains the results for two architectures.

\begin{figure*}
\begin{subfigure}[b]{0.15\textwidth}
  \includegraphics[trim=0 0 0 0, clip, height=1.9in]{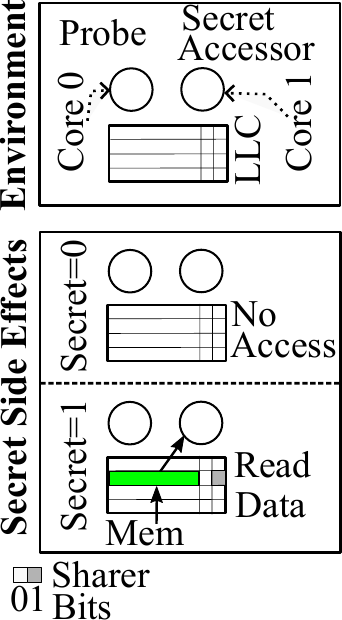}
  \caption{\textmd{Attack \\ Environment.}}
  \label{fig:attack_environment}
\end{subfigure}
\hspace{-0em}
\hfill
\begin{subfigure}[b]{0.25\textwidth}
  \includegraphics[trim=0 0 0 0, clip, height=1.9in]{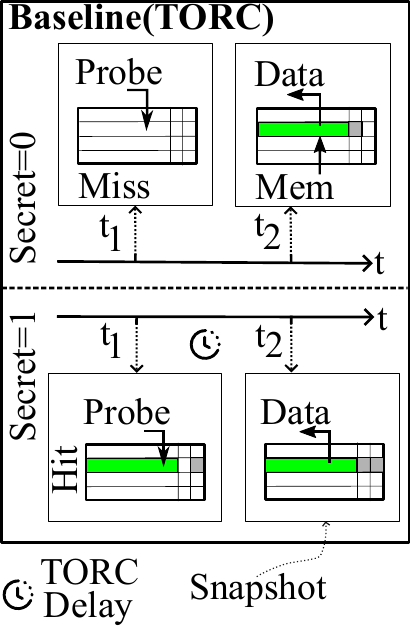}
  \caption{\textmd{Probe's Cache  \\ Snapshots For TORC.}}
  \label{fig:attack_probe_baseline}
\end{subfigure}
\hspace{-3.7em}
\hfill
\begin{subfigure}[b]{0.3\textwidth}
  \includegraphics[trim=0 0 0 0, clip, height=1.9in]{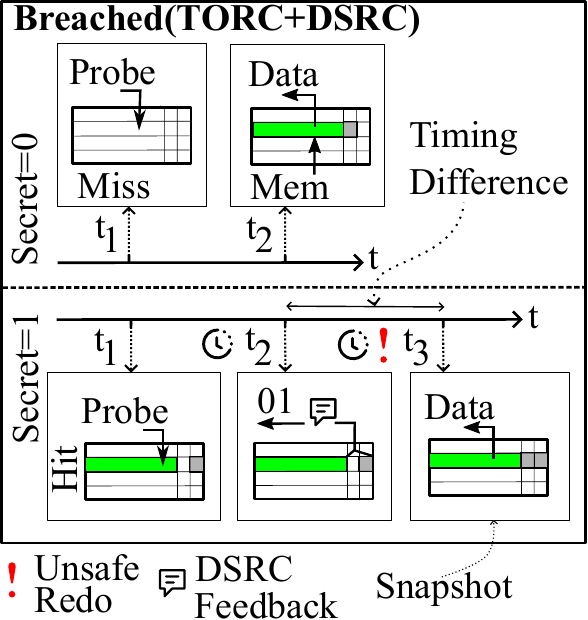}
 \caption{ \textnormal{Probe's Cache \\   Snapshots For TORC+DSRC.}} \label{fig:torc+dsrc}
\end{subfigure}
\hspace{-2.3em}
\hfill
\begin{subfigure}[b]{0.3\textwidth}
  \includegraphics[trim=0 0 0 0, clip, height=1.9in]{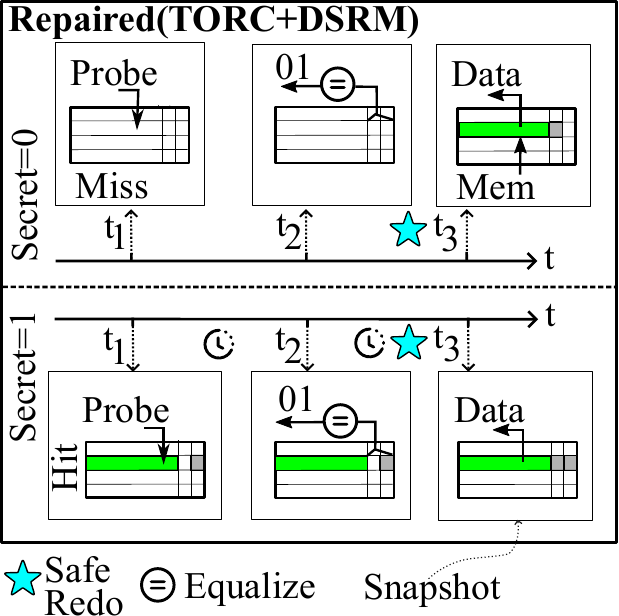}
  \caption{ \textnormal{Probe's Cache} \\ \textmd{Snapshots for TORC+DSRM.}}
  \label{fig:torc+dsrm}
\end{subfigure}
\\[-2ex]
\caption{ 
\textmd{Cache snapshots on a timeline while an attack probe runs. There are two scenarios, remote cache line absent (secret=0), and cache line present (secret=1), in three configurations. The remote cache line is represented as a green rectangle (data) with an attached sharer vector (two bits shown as squares) with the secret accessor core's bit set (gray) and the probe core's bit unset (white). The first configuration (Baseline) shows how TORC eliminates the probe's ability to detect a cache hit by delaying the timing of the hit at time $t_2$. 
 The second configuration (Breached) shows how the attacker abuses DSRC to induce an extra redo by exploiting the retrieval of remote cache line presence/absence at time $t_2$. The last configuration (Repaired), solves the problem by  forcing redos on both remote presence and absence, due to equalization at time $t_2$ (details  in \S\ref{sub:doing_attacks}).}}
\label{fig:ftm_attack_validation_delay}
\end{figure*}

\textbf{DSRC Defense.} The \emph{delaying speculative changes to remote cache lines} (DSRC) strategy does not permit unsafe speculative transitions to remote cache lines (e.g., \emph{E/M$\to$S} transitions). 
Instead, DSRC sends feedback to the core that a remote cache line is present, which causes a \emph{redo} operation if that load is confirmed to be non-speculative (e.g., when it reaches the head of the re-order buffer). DSRC is an important component in state-of-the-art speculative leakage defenses~\cite{ainsworth2020muontrap,ainsworth2021ghostminion}. 
We observe that a risk of DSRC is that it transmits remote presence information to the core, which could break the assumptions of other defenses.

\subsection{Need for Both TORC and DSRC}
\label{subsection:need_for_both}
TORC and DSRC mitigate different classes of attacks. DSRC cannot mitigate all the attacks that TORC mitigates because it only prevents speculative state changes and cannot prevent cache hits on non-speculatively inserted remote cache lines. On the other hand, TORC cannot prevent attacks that use a store-based probe where the attacker has induced the victim to make a speculative coherence state change in the former's address space. Hence, we need both TORC and DSRC to mitigate the two classes of attacks.

\textcolor{black}{
\textbf{Combining TORC and DSRC.} There are at least four ways to naively combine TORC and DSRC. When we combined TORC and DSRC, there can be two time points where TORC adds its delays. It can potentially add delays when a cache line is first accessed, or upon the redo access. Another alternative is to enforce a delay on both accesses, and the last option is to enforce no delay on any of the accesses. All these n\"{a}ive combination strategies result in a security issue, due to the net delay becoming much larger than a memory access time. We illustrate the security issue using the most straightforward combination strategy, i.e., employing a TORC delay on all remote cache hits.}

\section{Shield Bashing TORC Using DSRC}
\label{section:vulnerability}
Firstly, we present the environment in which TORC and DSRC are used together and the shield bash attack (see  \S\ref{subsection:environment} and \S\ref{sub:doing_attacks}).  Secondly, we illustrate how the DSRC defense can be abused to break the TORC defense (see \S\ref{subsection:break_torc}).  
Thirdly, we show that our improved defense, DSRM, eliminates the above attack, by preventing differential coherence state feedback into the core based on the presence or absence of a remote cache line (see \S\ref{subsection:mitigate}).
Fourthly, we discuss an SS-MESI approach to eliminate the attack (see \S\ref{subsection:non_dsrm}). 

\subsection{Attack Environment}
\label{subsection:environment}
We consider a setting where the program creating secret-dependent side-effects is located on one core (e.g.,  RSA, AES or a covert channel transmitter), and the probe is situated on another core. 
Each core has private cache levels and there is one shared last-level cache (LLC). The setting includes a standard MESI-like coherence protocol that uses sharer bit-vectors.
TORC and DSRC are both present as defensive capabilities in the hardware. 

\textbf{Initial Conditions.}
The upper part of Figure~\ref{fig:attack_environment} shows the probe on the core to the left (shown as a circle) and the secret accessor on the core to the right (also shown as a circle). A shared last-level cache used by both cores is  shown as a table.
The secret accessor corresponds to either a victim program (that is  spied on by an attacker), or to the transmitter of a covert channel.
The lower part of Figure \ref{fig:attack_environment}
shows the secret-dependent side-effect prior to deploying the probe. The secret accessor 
makes a decision, indicated by a secret value being zero or one, to not access or to access a memory location, respectively. This is similar to secret leakages in RSA or AES victims or to transmissions by covert channel transmitters. 
If the secret value is zero, it does not affect the cache state. If the secret value is one, it puts a remote cache line into the cache. 
The presence of a remote cache line is shown as the shaded green part of the cache. The sharer bit (right side) corresponding to the secret-accessor is set (gray). The sharer bit (left side) corresponding to the attacker is not set (white). If only one gray bit is set, it represents an \emph{exclusive} (E) state and if two bits are set, then it represents a \emph{shared} (S) state.

\subsection{Attack and Mitigations}
\label{subsection:attack_timelines}
We deploy a probe, which we refer to as \emph{shield bash of TORC using DSRC}. The key instruction in the probe is a load instruction that is on the right (taken) path but is issued to the cache while it is still in a speculative condition, as determined by DSRC (e.g., it is in a branch shadow, see \S\ref{section:attack}). Below, we examine how such a probe works in three possible configurations, a baseline TORC defense,  a TORC+DSRC defense and a TORC+DSRM defense.
\subsubsection{Baseline TORC Defense}
\label{sub:doing_attacks}

Figure~\ref{fig:attack_probe_baseline} shows the TORC defense operating on a typical microarchitecture.  
At time $t_1$, the probe attempts a load access into the cache (shown by a downwards pointing arrow).  
 There is a cache miss when secret is zero, otherwise, there is a cache hit.
The interval between $t_1$ and $t_2$ looks like a cache miss, either due to a main memory access (secret=0) or due to a TORC delay (secret=1). 
 At time $t_2$, data has been retrieved through the cache into the probe's core (shown by the upward pointing arrows).
 Thus, baseline TORC causes the attacker's timing measurement to  resemble a cache miss, regardless of the secret value.
 
\subsubsection{Breaking TORC by Abusing DSRC}
\label{subsection:break_torc}
Figure~\ref{fig:torc+dsrc} shows how DSRC is used to break the TORC defense. The attack exploits the microarchitectural defense assumption violation that only data is sent to the core upon a cache hit and not other information about the cache hit.
The events up to time $t_2$ remain the same as the baseline. 

At time $t_2$, if the remote cache line is absent (secret = 0), data is retrieved from the main memory and transferred to the core through the cache. The load instruction has completed its cache interactions. If the remote cache line is present (secret = 1), the remote cache line's presence is retrieved by the DSRC feedback logic in the cache (shown as a chat icon).  This is done by reading the sharer bits from the cache line (01) and returning them to the core after a TORC delay. When the load instruction is no longer speculative, the core triggers a redo based on DSRC's feedback about a cache hit on a remote cache line. The redo gets the latest data and updates the cache coherence state. The redo is signified by a red exclamation mark on the timeline. Due to a TORC delay, the redo takes time resembling a cache miss (shown by a clock icon on the corresponding timeline). 
At time $t_3$, data from the redo has reached the core, thus completing the load instruction's interaction with the cache.

The probe observes a timing difference of one cache miss time when the secret is zero versus two cache miss times when the secret is one.
Thus, the secret is leaked, breaking the TORC defense, which was otherwise effective in the absence of DSRC.

\subsubsection{DSRM Eliminates the Abuse}
\label{subsection:mitigate}
Figure~\ref{fig:torc+dsrm} demonstrates that using DSRM, instead of DSRC, together with TORC (on MESI) eliminates the shield bash attack. 
The events until time $t_2$ are the same as the baseline. At time $t_2$, DSRM sends the same feedback of `01' to the core, i.e., it equalizes the feedback regardless of actual presence or absence of the remote cache line (indicated by the equals symbol).
Once it confirms that the load instruction is on the correct path, the core triggers a redo of the load. The action of carrying out the redo non-selectively is marked with a light blue star symbol in both situations. At time $t_3$, the data from the cache miss has returned to the core (secret=0) or alternately, the TORC-delayed data from the remote cache line has returned to the core (secret=1). The probe observes no timing difference.
Hence, DSRM eliminates the shield bash attack.

\textbf{Performance Optimization.} \textcolor{black} {We remove TORC delays in the time interval $t_1$ to $t_2$. The total time spent resembles an LLC access and a cache miss (instead of two cache misses), regardless of the presence or absence of the remote cache line. Hence, the above optimization maintains the security properties of DSRM.}

\subsubsection{Using SS-MESI With TORC}
\label{subsection:non_dsrm}
\textcolor{black} {
In the SS-MESI protocol (similar to MESI except that we always start with S state on load miss), by contrast, the remote cache lines inserted by the secret-accessor are always in an S state. DSRC never triggers redo operations on an S state cache line, eliminating the shield-bash attack. The overhead of DSRM under the MESI protocol is related to the number of speculative cache misses but the overhead of SS-MESI is dependent on the number of upgrades from S to M.}

\begin{tcolorbox}[colback=gray!5!white,colframe=gray!75!black,title=Takeaway 1]
    Our attack breaks TORC by abusing the redos introduced by DSRC. DSRM and SS-MESI eliminate the attack by always causing redos, or by never causing redos, regardless of the secret.
\end{tcolorbox}

\section{Shield Bash: Implementation and Mitigations}
\label{section:attack}

In this section, we first present the hardware support necessary for integrating TORC and DSRC defenses onto a standard out-of-order core with shared LLC (see \S\ref{subsection:implementation}). This creates an MDAV that \textcolor{black}{facilitates} the shield bash attack. We also implement our mitigations, DSRM and SS-MESI.
We then present our implementation of a probe that can trigger redos in the core based on presence or absence of remote cache lines, using these hardware supports (see \S\ref{subsection:attack_impl}). 
Although this probe works for both speculation protection decision models, BranchShadow and ROB-Head, we focus on the BranchShadow model, as it is a weaker model and it is widely used~\cite{yan2018invisispec, aimoniotis2023recon}.

 \subsection{Implementing the Attack Environment}
\label{subsection:implementation}
We build our attack environment on top of a cache coherence protocol is an inclusive three-level cache using MESI protocol similar to GEM5's inclusive 3-level cache protocol~\cite{ruby}. We integrate TORC and DSRC support onto standard out-of-order cores similar to a standard O3~\cite{lowe2020gem5} model with speculation support (e.g, branch and load reordering speculation). 

 \subsubsection{TORC} 
 When a remote cache line hit in the LLC needs to be delayed, a main memory access is sent for creating delay. We buffer the response from the cache hit in a private buffer near the requestor core until the delay access returns and releases it to the core~\cite{ramkrishnan2020first}.

\subsubsection{DSRC}
 Hardware support is integrated to enable the  identification of instructions that can be affected by speculative attacks, such as the \emph{E/M$\to$S} attack variants (see \S\ref{subsection:speculative_leakages}).  There are two commonly used \emph{speculation protection decision models} (SPDM): BranchShadow~\cite{yan2018invisispec,yu2019speculative,aimoniotis2023recon} and ROB-Head~\cite{yan2018invisispec,ainsworth2021ghostminion,ainsworth2020muontrap}. In \emph{BranchShadow}, the load is issued speculatively when it is in a branch shadow, i.e., there is an older branch in the ROB which has not yet resolved. In the \emph{ROB-Head} model, all loads are issued speculatively, unless the load was at the head of the ROB at the time of issue. 
A speculation protection flag that represents the result of the speculation protection decision is added to a load 
request (by checking the ROB) before it is sent to the cache. A zero-valued flag indicates that it is not a protected load and hence, does not need to trigger DSRC feedback. A flag value of one indicates that it could be an unsafe speculative load and needs to be protected. It does need to trigger DSRC feedback. GETS\footnote{A GETS request in a MESI protocol refers to a read issued by a cache controller to the next cache level upon a read miss. It is to get data and to update coherence state.} requests from L1 to L2, and from L2 to L3, also contain this flag. 

\emph{DSRC Cache Feedback.} The flag is finally used at L3, where cache hits on the L3 cache lines are checked for remote E/M coherence state. In case the flag is set, an additional check is carried out. If remote E/M coherence state is found, then it is sent back to the accessing core, as a REMOTE-EM response message. If the flag is not set, the regular MESI protocol is applied. If REMOTE-EM is returned to the processor, it reissues the load if declared safe by the speculative protection decision model.

\subsubsection{Mitigations}
The equalization operation of \textbf{DSRM} is implemented by sending a REMOTE-EM response message back to the private caches on both cache hit of a remote cache line and cache miss. The equalization operation happens only if the GETS had its speculation protection flag set to one. The other aspects of the MESI protocol, such as coherence states, are unchanged. 
In \textbf{SS-MESI}, we add a control flag to the LLC which directs all load misses in the LLC to start in the S state instead of E state. 

\subsection{Implementing the Attack Code}
\label{subsection:attack_impl}
Firstly, we discuss three key instructions that are a part of the LRBS probe. Secondly, we  use a timing diagram to explain how the LRBS probe functions (see \S\ref{subsub:timing_diagram_miss}).  
Lastly, we describe an \texttt{x86\_64} implementation of this probe (\S\ref{subsection:asm_description}).

\subsubsection{Attack Code: Key Instructions}

Firstly, we have the Load-Before-Branch (LBB) instruction, which has a cache miss during execution. This amplifes speculation effects (i.e, it creates a large speculation window).
LBB accesses occur to a cache line address that is not touched by the secret accessor and is under the control of the probe only.
Secondly, we have a branch instruction, which is dependent on the result of this load instruction, and which always resolves as not taken. This branch is the source of speculation within the probe. Lastly, we have the Load-After-Branch (LAB) instruction,  which is on the not-taken path. This LAB instruction exploits the speculation to trigger redo operations. The LAB load's cache line address corresponds to the cache line address that is touched by the secret accessor.


\textbf{Training the Probe. }We first train the branch predictor by running the probe several times. 
The branch predictor is trained as `not taken', so that the core executes the LAB instruction speculatively before branch resolution. After training has completed, it is ready to detect the side-effects left behind by the secret accessor program.

\begin{figure}
  \includegraphics[width=\linewidth]{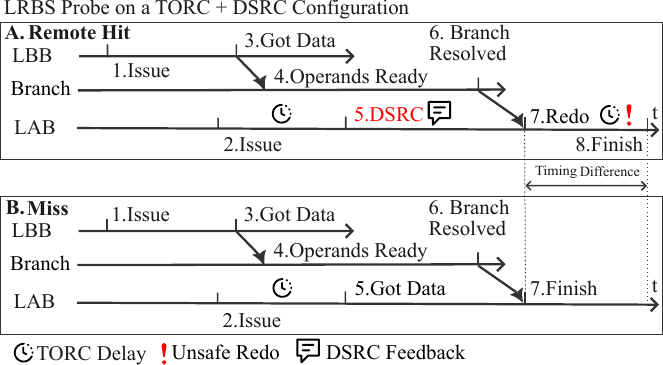}
  \caption{\textmd{A timeline showing how the MDAV caused by sending coherence information to the core is exploited using the LRBS probe (see \S\ref{section:attack}) on a TORC + DSRC configuration. In case A, the timeline of the probe is shown, when a remote cache line in the E state is present in the LLC (slower due to redo). In case B, the timeline is shown when there is a cache miss (faster).}}
  \label{fig:ftm_breaking_receiver}
\end{figure}

\subsubsection{Creating Timing Differences in  TORC + DSRC}
\label{subsub:timing_diagram_miss}

The probe's goal is to trigger redos conditionally according to the presence or absence of a remote cache line. Figure~\ref{fig:ftm_breaking_receiver} shows an example timing diagram where the above requirement is met. There are two cases, one is for a cache hit  on a remote cache line (case A) and the other is for a cache miss (case B).

\textbf{Remote Cache Line Hit.}
We first discuss the event timeline for a remote cache line hit scenario. 
The LBB and LAB instructions issue load accesses into the cache. The LBB issues before the LAB because it is earlier in the program order and thus gets dispatched to the execution stage first (time-point \emph{1}). Due to its earlier issue, the LBB load obtains data from the cache first (time-point \emph{3}). The branch instruction's resolution condition has a data dependency on the speculative load access (shown by an arrow from time-point \emph{3} to time-point \emph{4}).  Thus, the LBB load miss is crucial in extending the branch resolution until a later time point, as opposed to a much earlier resolution if there were an LBB cache hit. 

Meanwhile, the LAB load gets issued speculatively (time-point \emph{2}), for performance reasons, while it is still in the branch shadow. Soon after, the branch condition is available to the branch instruction (time-point \emph{4}). Sometime after that, the branch is resolved (time-point \emph{6}). The LAB load obtains a response (remote E state) from the cache, after the cache has applied a TORC delay on its remote cache line hit (time-point \emph{5}). However, the LAB load cannot commit yet because it was issued speculatively into the cache and thus it did not receive any data upon a remote cache line hit. Hence, the LAB instruction waits in the pipeline until the branch is resolved (time-point \emph{6}). The core then re-issues the LAB load to the cache (time-point \emph{7}), where the load sees a TORC delay and finally returns the data to the core (time-point \emph{8}).

Time-points $5$ and $7$ are crucial parts of the exploitation of the vulnerability. At time-point $5$ the DSRC feedback occurs and at time-point $7$ the unsafe redo operation (indicated by a red exclamation mark) occurs. The DSRC feedback is abused by the attacker to trigger the redo, which ultimately causes a secret-dependent timing difference.

\textbf{Cache Miss.}
\label{subsub:timing_hit}
During a cache miss (Figure~\ref{fig:ftm_breaking_receiver}-B),  a similar event sequence occurs, except for two key differences in the LAB's timeline. The first difference is that, at time-point 5, data is returned to the core but not a coherence state. Second, there is no redo step for the LAB, so it finishes earlier.
The difference in timing measurement between a cache miss and a remote cache hit is indicated using a dotted pointer.

 \subsubsection{Realizing Shield Bash On x86\_64}
 \label{subsection:asm_description}
 Listing~\ref{lst:slow_pipe_receiver} shows an implementation of the LRBS probe. In this probe, the LBB is on line 8 and the LAB is on line 11. The LBB loads data into the \texttt{\%r12d} register. The branch instruction is on line 10, using a \texttt{jne} that is dependent on \texttt{\%r12d}. The timer measurement starts on line 5, prior to the LBB, and ends on line 14 after the branch completes its jump to line 12. Load fences serialize loads before and after the timer-start event. Load fences are similarly applied to the timer-end event. Finally, line 15 calculates the timing difference and lines 16 and 17 clear the cache lines associated with the LAB and LBB. Line 18 indicates that the timing result is recorded in the \texttt{\%eax} register. Line 19 indicates the two registers that contain the addresses of LBB and LAB loads. 

\subsubsection{Applying Mitigations.}
\label{apply_mitigations}
One possible strategy to improve resilience against LRBS is to avoid TORC delays on speculative accesses. This may potentially reduce the delays on the LRBS probe by making the LAB return coherence information faster. However, it does not completely mitigate the shield bash attack because the LAB still needs to redo with a TORC delay once the LAB is confirmed as non-speculative.
On the other hand, DSRM and SS-MESI mitigations are effective by always creating redos or by eliminating redos of the LAB from the LRBS probe's execution. 

\begin{lstlisting} [language=C, caption={ \textmd{An implementation of the LRBS probe. The functionality of the code, which is to trigger secret-dependent redo operations, is explained in \S\ref{subsection:asm_description}.} },  label={lst:slow_pipe_receiver}, captionpos=b, float]
//LRBS Probe
1 asm __volatile__ (
2 " xorq %%r12, %%r12\n" //Initialize %r12 to zero
3 " mfence \n"  //Serialize older mem instructs 
4 " lfence \n"  //Serialize loads
5 " rdtsc \n"   //Start timer and store in %eax
6 " lfence \n"  //Serialize loads
7 " movl %%eax, %%esi \n" //Save timer into %esi
<@\textbf{\texttt{\textcolor{black}{8 \hspace{0.7mm}"\hspace{0.6mm}  movl (\%2), \%\%r12d~~\textbackslash n" }}}@> //LBB load
9 " testl %%r12d, %%r12d\n" //Set equals flag 
<@\textbf{\texttt{\textcolor{black}{10 \hspace{0.0mm}"\hspace{0.0mm} jne \%=f\textbackslash n"}}}@> //Branch makes speculative shadow
<@\textbf{\texttt{\textcolor{black}{11 \hspace{0.0mm}"\hspace{0.0mm} movl (\%1), \%\%eax~~\textbackslash n"  }}}@> //LAB load (redo)
12 " %=:\n" //Jump target for the jne on line 10
13 " lfence \n" //Serialize loads
14 " rdtsc \n" //End timer stored in %eax
15 " subl %%esi, %%eax \n" //<@$\Delta T$@> = %esi - %eax
16 " clflush 0(%1) \n" //Flush LAB line address 
17 " clflush 0(%2) \n" //Flush LBB line address
18 : "=a" (time) //Output C variables gets <@$\Delta T$@>
19 : "c" (LAB), "r" (LBB) //C variables
20 : "%esi", "%edx", "%r12"); //Clobbered regs
\end{lstlisting}

\section{Attack Simulations of the LRBS Probe}
\label{subsection:attack_experiments}

The common GEM5 simulator configuration is shown in  Table~\ref{tab:Simulation_Configuration}. We discuss the attack simulation configurations and the attack simulation results (see \S\ref{subsub:attack_result_summary}).

\subsection{Attack Simulation Configurations} 
We implement five configurations  in GEM5 for each attack simulation: $C_1$, $C_2$, $C_3$, $C_4$ and $C_5$. 
$C_1$ corresponds to an insecure cache configuration. $C_2$ corresponds to a TORC implementation. $C_3$ corresponds to a TORC + DSRC configuration (TORC is only applied to non-speculative accesses in an attempt to reduce LRBS execution time, see \S\ref{apply_mitigations}).
$C_4$ corresponds to a TORC + DSRM configuration (the optimized version with less TORC delays, see \S\ref{subsection:mitigate}) and $C_5$ corresponds to a T + DC + SS-MESI configuration, which is a combination of TORC, DSRC and SS-MESI.
The attack transmitter creates no cache side-effect if the secret value to be transmitted is zero. Otherwise (secret value one), it places a remote cache line into the LLC. We carried out two measurements per attack simulation, one for a secret value of zero transmitted by the transmitter, and one for a secret value of one transmitted by the transmitter (totally 10 experiments). We ran each experiment at least 100 times and recorded the median timing result.

\subsection{Attack Result Summary}
\label{subsub:attack_result_summary}

Table~\ref{tab:attack_results} (for the attack results) has one row for each of the configurations $C_1$, $C_2$, $C_3$,  $C_4$ and $C_5$. The different entries in each column indicate the timing measurements made for each attack. A difference between the timing columns indicates a successful attack.

\begin{table}
\small
\begin{tabular} { | p{3.9cm} | p{1.6cm} | p{1.6cm} |
}

\hline
\multicolumn{3}{|c|}{Attack Simulations Using LRBS} \\
\hline
\textbf{Defense Config} &  \textbf{Secret=0}  & \textbf{Secret=1}   \\
\hline

 \multirow{1}{*}{1. Insecure ($C_1$)} &   \text{205}  &  \text{199}  \\
\hline

\multirow{1}{*}{2. TORC ($C_2$)} &  \text{205}  &   \text{205} \\

\hline

\multirow{1}{*}{3. TORC + DSRC ($C_3$)} &  \text{205}  &  \text{364}  \\
\hline

\multirow{1}{*}{4. TORC + DSRM ($C_4$)} &    \text{364}  &    \text{364}  \\

\hline

\multirow{1}{*}{5. T + DC + SS-MESI ($C_5$)} &    \text{205} &   \text{205} \\
\hline

\end{tabular}
\\[2ex]
\caption{
 \textmd{$C_1$,  $C_2$,  $C_3$, $C_4$ and $C_5$ correspond to Insecure, TORC, TORC + DSRC, TORC + DSRM and T + DC + SS-MESI, respectively. $C_3$ does not have the
attack resilience we would expect from the union of TORC and DSRC defenses. DSRM or SS-MESI restore resilience against the LRBS probe.}}
\label{tab:attack_results}
\end{table}

\textbf{Insecure ($C_1$).}
When the secret is zero then there are cache misses on both LAB and LBB. However, the two cache misses mostly overlap each other because the insecure configuration allows all accesses to create side-effects without timing restriction. We observed a timing of 205 cycles. 
When the secret is one, the LAB load has a cache hit on the cache line. However, the access to the load which delays the branch (LBB) will still miss. Hence, the total time is expected to be a bit less than the case where the secret is zero. Our observation of 199 cycles is consistent with the above expectation.

\begin{table}[t]
\small
\begin{tabular}{ |p{1.9cm}|p{6cm} |
 }
 \hline
 \multicolumn{2}{|c|}{\textbf{GEM5:} Simulation Configuration} \\
 \hline
 \textbf{Component} & \textbf{Configuration} \\
 \hline
 Core &    3GHz, OOO core, 192-entry ROB, 32 entry LQ, 32 entry SQ,  fetch/commit/squash width of 8, tournament BP, SMT disabled \\
 \hline
 L1D Cache & 32 KB, 8-way associative, 2 cycle latency, Pseudo-LRU replacement \\
 \hline
 L1I Cache & 32 KB, 8-way associative, 2 cycle latency, Pseudo-LRU \\
 \hline
 L2 Cache & 256 KB, 8-way associative, 16 cycle latency, Pseudo-LRU replacement \\
 \hline
 L3 Cache &  2MB slices, 16-way associative, 40 cycle latency, Pseudo-LRU replacement \\
 \hline
 Interconnect & 4x2 Mesh\_XY, 16-byte link width, 1 cycle link latency, 1 cycle router latency \\
 \hline
 Coherence & MESI\_Three\_Level Protocol \\
 \hline
 Main Memory & DDR4\_2400\_8x8, 140 cycles latency  \\
 \hline
\end{tabular}
\\[2ex]
\caption{\textmd{Simulation configurations using GEM5 v23. We use the above configuration for demonstrating the security issue due to the microarchitectural defense assumption violation caused by the DSRC defense with respect to TORC.}}
\label{tab:Simulation_Configuration}
\end{table}

\textbf{TORC ($C_2$).}
 When the secret value is zero, both the LAB and the LBB load miss on their respective cache line addresses. When the secret value is one, the LBB has a cache miss and the LAB has a cache hit. TORC delays the cache hit on LBB and makes it look like a cache miss. Hence, timings are similar in both cases (205 cycles), which keeps the secret protected.

When the secret is zero, the measurement of the LRBS probe is 205 cycles due to two cache misses which overlap partially, one for the LBB and the other for the LAB. When the secret is one, the measurement of the LRBS probe is 364 cycles, which is much larger than the time recorded when the secret value was zero. The extra time is due to a redo event that occurs on the LAB load.

\textbf{TORC+DSRM ($C_4$).}
For both secret value zero and secret value one, the timing observed is 364 cycles. 
For secret value one, the timeline of the probe is similar to DSRC. For secret value zero, the LAB undergoes a dummy coherence feedback (equalization) and a cache miss, which makes the timing similar to the case where the secret is zero (364 cycles).

\textbf{T+DC+SS-MESI ($C_5$).}
\textcolor{black} {For both secret value zero and secret value one, the timing observed is 205 cycles, due to a TORC delay occurring on the LBB and either a cache miss or a TORC delay occurring on the LAB.}

\begin{tcolorbox}[colback=gray!5!white,colframe=gray!75!black,title=Takeaway 2]
TORC+DSRM (Repaired) and SS-MESI  (Repaired) are resilient against the LRBS attack probe whereas TORC+DSRC (Breached) is not resilient against it.
\end{tcolorbox}

\textcolor{black}{
\subsection{Transmission Rate on a Real Machine}
We  measure the timings of the above  covert channel on a real machine (Xeon E5-2699 v3). 
Our epoch size is 1 million cycles. The error rate is 0.3\% as measured across 196608 bit transmissions. The effective transmission rate is 6KB/s. In this experiment, 1/0 are transmited as accessing/not accessing a particular cache line. We simulate a long cache hit time on remote cache lines hits by invoking a delay loop upon detecting remote cache hits. For error correction, we conservatively transmitted each bit 16 times.}

\textcolor{black}{
\textbf{Security Analysis of the Covert Channel.}
We model the covert channel abstractly as consisting of a transmitter and a receiver, operating on a common shared cache. There is a group of cache line addresses, $\mathcal{A}$, which can fit in the cache together. The transmitter accesses a subset of this group of addresses and populates the cache. In the second step, the receiver carries out a sequence of speculative right-path memory accesses using a receiver probe. The probe accesses a subset of the addresses $\mathcal{A}$. Under these conditions, the required security property is that, the receiver has a constant total timing, regardless of the initial cache state that was created by the transmitter. The transmitter then flushes the cache and then repeats both the steps.}
\textcolor{black}{ 
SS-MESI straightforwardly enables the security property to be satisfied. Each access to a remote cache always takes a cache miss amount of time. For each step, the total time taken by the receiver is always the sum of the total number of memory accesses, or $T_M * N$, where $T_M$ is the memory access time and $N$ is the number of addresses it accesses. For each step, the total time taken by optimized version of DSRM is $(T_C + T_M)*N$, where $T_C$ is the LLC access time, $T_M$ is the memory access time and $N$ is the total number of addresses accessed. 
}

\section{Performance Costs Of Fixing Shield Bash}
\label{section:performance}
We simulate defense configurations TORC + DSRC ($C_3$) and TORC + DSRM ($C_4$). We use both the BranchShadow and ROB-Head speculation protection decision models. We also simulate the T + DC + SS-MESI ($C_5$) configuration, which is also resilient against the new attack).
The common GEM5 processor and cache configurations are the same as we summarized in Table~\ref{tab:Simulation_Configuration}.
We simulate two classes of workloads, namely, PARSEC workloads and SPECrate 2017 workloads, which stress the processor microarchitecture in multi-threaded, single-threaded and multiprogramming scenarios. We first present the PARSEC performance results (see \S\ref{subsection:multithreaded}) and then, we present the SPECrate 2017 performance results (see \S\ref{subsection:spec}).

\emph{Hardware Costs.}
Our implementation of DSRM, SS-MESI and DSRC, primarily modifies the the hardware's control logic, which takes up a relatively small amount of resources. The TORC buffer entries contain address, data and coherence state of delayed responses to the core. For 48 bits of physical address,  8 bytes of data  and 1-bit of coherence information, in a 32-entry buffer (max outstanding load requests), this is about 0.4KB of storage per core. 

\subsection{PARSEC Evaluation (Multithreaded Workloads)}
\label{subsection:multithreaded}
The PARSEC workloads represent a range of multi-threaded access patterns, with a significant amount of shared data and code between different threads. 
We use a simulation methodology similar to that used by Singh et al.~\cite{singh2012end} on an eight core processor configuration. Checkpoints are taken on PARSEC benchmarks every 150 million cycles starting from the region of interest, until the end of the workload (\emph{simlarge} size). Then, we simulate enough instructions so there are at least 10 million stores on each core on each checkpoint, for a total of at least 80 million stores. The microarchitectural state is warmed up by simulating at least 100,000 stores and only collect statistics after that point.  We simulate six sample checkpoints, each one five times. 
The performance metric is averaged over all simulations in each configuration. Thus, we simulate enough instructions to include at least 2.4 billion stores for each PARSEC benchmark.

\textbf{Overhead Measurement.}
In sampling multithreaded workloads, we cannot guarantee that we simulate exactly the same amount of work for each configuration's simulation, so direct comparisons between the cycles executed during two simulations can be misleading. Therefore, for our overhead measurement, we measure the total number of cycles for each configuration. The other two steps are to measure the total number of TORC delays and DSRC/DSRM redos, and to take the ratio between the overhead cycles and the non-overhead cycles.
 We also present other supporting metrics such as the fraction of cache accesses that reach the LLC and  the fraction of LLC accesses that trigger redos, to better understand our observations. 

\begin{figure}
\begin{subfigure}{\linewidth}
  \includegraphics[trim=0 93 0 0, clip, width=\linewidth]{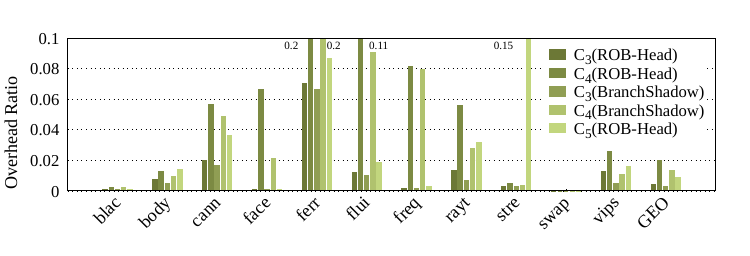}
  \\[-5.5ex]
  \caption{ \textmd{The performance overhead results for the PARSEC benchmarks are presented above. The geometric mean of the  performance overheads are less than 2.8\% for all configurations.}}
  \label{fig:parsec_results_ipc}
\end{subfigure}
\\[-0ex]

\begin{subfigure}{\linewidth}
  \includegraphics[trim=0 90 0 0, clip, width=\linewidth]{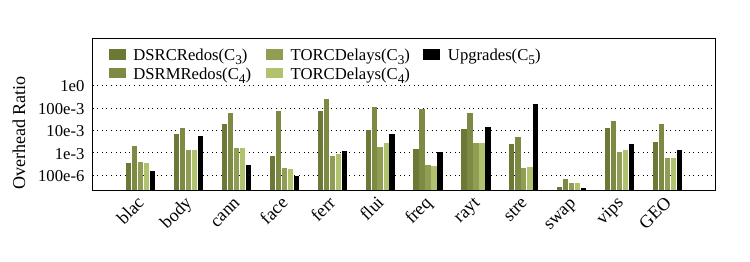}
  \\[-5ex]
  \caption{\textmd{ Cycle overhead ratio due to different sources, namely, DSRC redos, DSRM redos and TORC delays on the ROB-Head model. The trends and values are similar for BranchShadow model. The last bar shows the ratio of extra cycles to non-overhead cycles, in upgrading S to M state for the T + DC + SS-MESI configuration in the ROB-Head speculation protection decision model.} }
  \label{fig:overhead_breakdown_ROBHead}
\end{subfigure}
\\[-2ex]
\caption{\textmd{PARSEC simulation results (see \S\ref{section:performance}).}}
\label{fig:parsec_results}
\end{figure}

\textbf{Performance Results.} 
Figure~\ref{fig:parsec_results_ipc} shows that performance overheads for the five simulated configurations. They geometric means of the overheads for the five configurations, $C_3$ (ROB-Head), $C_4$(ROB-Head), $C_3$(BranchShadow), $C_4$(BranchShadow) and $C_5$(ROB-Head) are 0.0038, 0.019, 0.0028, 0.013 and 0.0080. Thus, the total overheads over all configurations are 2.8\% or less.
The geovariance around the geometric means are all less than 1\%. 

\textcolor{black}{
 Figure~\ref{fig:overhead_breakdown_ROBHead} shows more specific overheads for the ROB-Head speculation protection model (which is generally higher than the BranchShadow model).} For the $C_3$  configuration, the TORC overheads are 0.00051 and the DSRC overheads are 0.0029. For the $C_4$ configuration, the TORC overheads are 0.00053 and the DSRM overheads are 0.017. For the $C_5$ configuration, the overheads caused by S$\to$M are 0.0012.

\begin{figure}
\begin{subfigure}{\linewidth}
  \includegraphics[trim=0 90 0 0, clip, width=\linewidth]{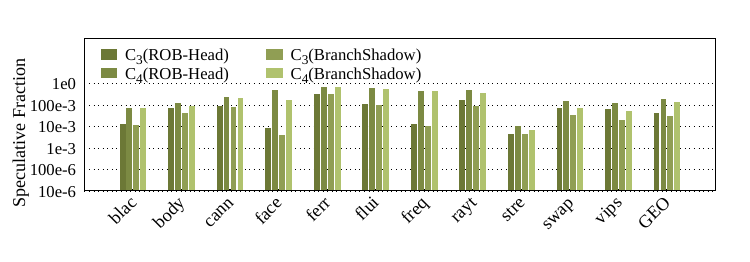}
  \\[-5ex]
  \caption{\textmd{The fraction of LLC accesses that trigger redos due to DSRC or DSRM. Higher this speculative fraction, the greater the amount of overheads due to DSRC and DSRM redos.}}
  \label{fig:parsec_dsrm_fraction}
\end{subfigure}

\begin{subfigure}{\linewidth}
  \includegraphics[trim=0 90 0 0, clip, width=\linewidth]{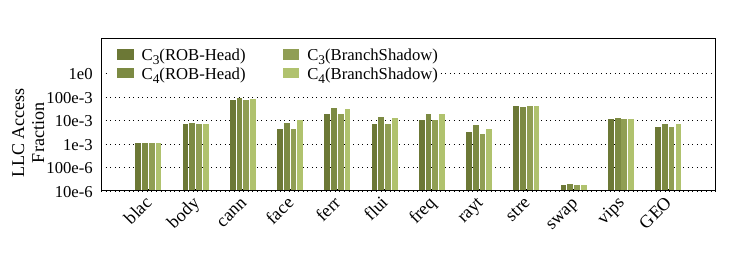}
  \\[-5ex]
  \caption{\textmd{The number of LLC accesses as a fraction of the total number of cache accesses. A greater LLC (access) fraction creates a higher chance of delays and overheads.}}
  \label{fig:llc_access_fraction}
\end{subfigure}
\\[-2ex]
\caption{\textmd{Supporting data for understanding PARSEC cycle overheads  (see \S\ref{section:performance}).}}
\label{fig:parsec_support}
\end{figure}

To support and substantiate the above results, we have Figure~\ref{fig:parsec_dsrm_fraction}, which  shows the fraction of the total number of cache accesses into the LLC, that cause DSRC or DSRM redos. The fraction is shown for both the ROB-Head and BranchShadow speculation protection decision models. For the ROB-Head model, the average speculation fraction for DSRC in $C_3$ is 0.041 and for DSRM in $C_4$, it is 0.18. For BranchShadow, the average speculation fraction for DSRC redos in $C_3$ is lower (0.027) and it is higher for DSRM in $C_4$ (0.13).

\textcolor{black}
{Figure~\ref{fig:llc_access_fraction} shows the number of LLC accesses as a fraction of the total number of cache accesses. This value is similar across configurations in most cases. The values for the configurations $C_3$ (ROB-Head), $C_4$(ROB-Head), $C_3$(BranchShadow), $C_4$(BranchShadow) and $C_5$(ROB-Head) are 0.0049, 0.0068, 0.0048, 0.0064 and 0.0057.}

\emph{Performance Trend.} 
The configurations corresponding to  $C_4$ have a higher overhead mainly because DSRM causes additional redos that DSRC does not cause, on cache misses. The overheads for the BranchShadow speculative protection decision model are less than the ROB-Head speculative protection decision model because of a smaller fraction of LLC accesses being protected by redos.

\textcolor{black} { \emph{Edge Case.} The \texttt{streamcluster} workload has a higher fraction of demand accesses which cause S$\to$M (10\%). The DSRM repair has significantly lower overhead cycles (2\%).}

\subsection{SPECrate 2017 Single Core}
\label{subsection:spec}
\subsubsection{Single Core} We evaluate a single benchmark on a single core to observe the performance impact of using DSRM instead of using DSRC. We used a methodology of fast-forwarding 20 billion cycles to skip initialization~\cite{beckmann2013jigsaw}. Then, each benchmark executed for at least 250 million instruction. We recorded the IPC of the core over that duration. As a performance comparison baseline, we also simulate the $C_1$ (insecure) configuration, which corresponds to an insecure system that does not have hardware defenses against side-channels. This is in addition to the key configurations $C_3$(ROB-Head), $C_3$(BranchShadow), $C_4$(ROB-Head) and $C_4$(BranchShadow) and $C_5$ (ROB-Head).

\begin{figure*}[t]
\begin{subfigure}{\linewidth}
  \includegraphics[trim=0 92 0 40, clip, width=\linewidth]{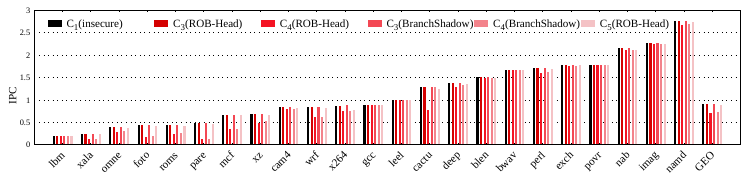}
  \caption{ \textmd{The performance results for single core SPEC simulations. The benchmark results are sorted in ascending order of IPC (insecure).}}
  \label{fig:spec_results_single_core}
\end{subfigure}
\\[1.5ex]
\begin{subfigure}{\linewidth}
  \includegraphics[trim=0 85 0 40, clip, width=\linewidth]{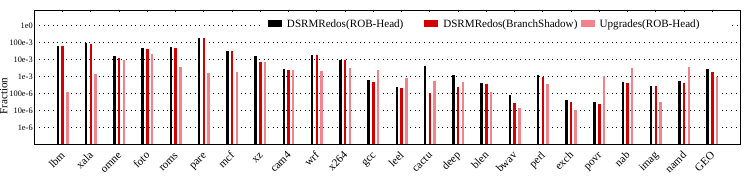}
  \\[-6.0ex]
  \caption{ \textmd{The fraction of memory accesses that are DSRM redos and stores carrying out upgrades from S to M.}}
\label{fig:spec_overheads}
\end{subfigure}
\\[-1.0ex]
\caption{\textmd{SPECrate 2017 single core performance results, the DSRM fraction and the upgrade fraction (\S\ref{subsection:spec}).}}
\label{fig:spec_single_core}
\end{figure*}

We record the instructions-per-cycle (IPC) metric for each workload (see Figure~\ref{fig:spec_results_single_core}). 
 The IPC for $C_1$ (insecure), $C_3$(ROB-Head), $C_3$(BranchShadow), $C_4$(ROB-Head) and $C_4$ (BranchShadow) and $C_5$ (ROB-Head) are
0.91, 0.91, 0.69, 0.91, 0.72 and 0.88.
The drop in performance for TORC+DSRM is mainly due to the significant number of cache misses that are incurred in many SPEC benchmarks, which are more expensive due to the extra redo operations. Thus, the average performance overheads are less than 24\% across both $C_4$ configurations. On average, the overheads for the T + DC + SS-MESI are about 2.8\%.
Figure~\ref{fig:spec_overheads} presents the fraction of LLC accesses which are DSRM redo operations. The fraction of demand accesses that trigger a DSRM redo are 0.0027 and 0.0018 for $C_4$(ROB-Head) and $C_4$(BranchShadow). The fraction of accesses which cause S$\to$M are only about 0.001 on average.

\emph{Performance Trends.} 
A higher IPC generally also has a lower DSRM redo fraction. This is evident from the shorter IPC bars on the left half of Figure~\ref{fig:spec_results_single_core} compared to its right half, and correspondingly, taller DSRM redo fraction bars in the left half of Figure~\ref{fig:spec_overheads} compared to its right half.  
A higher number of DSRM redos generally lowers the IPC significantly compared to the insecure baseline. This can be seen in that the height difference at lower IPCs between $C_1$ and $C_4$ is usually  larger, compared to the height difference at higher IPCs.

\emph{Edge Cases.} 
The $C_5$ (ROB-Head) IPC for \texttt{x264} is 0.76 versus an insecure IPC of 0.85. The upgrade fraction is on the higher side, at 0.39\% compared to the average of 0.085\%.
\textcolor{black} {For \texttt{blender} and \texttt{imagick}, the number of DSRM redos is on the lower side,  which helps them to have slightly better performance (0.2\% and 0.3\%) than the SS-MESI configuration.}

\captionsetup{%
   justification=centering,
   labelfont=bf,
  singlelinecheck=off
}

\captionsetup[subfigure]{%
   justification=centering,
   labelfont=bf,
  singlelinecheck=off
}

\subsection{SPECrate 2017 Multiprogramming Workloads}
\label{subsection:spec_multiprogramming}

\textbf{Simulation Methodology.}
We use a standard simulation methodology, where eight cores execute eight different SPEC benchmarks. Each workload is fast-forwarded by 10 seconds to ensure that the initialization phase of the workloads is completed.
A microarchitectural simulation of at least 250 million instructions is carried out for each workload.
We use a bound-and-weave model~\cite{sanchez2013zsim} to parallelize the GEM5 simulation. A default phase interval of 3 cycles (999 ticks) is used.

\textbf{Workload Mixes.} We classify the benchmarks into 35 types based on a standard approach~\cite{beckmann2013jigsaw}. The benchmarks are classified in four different ways. Either cache friendly, cache agnostic, cache fitting or streaming (based on the MPKI of the workloads). We then mix-and-match the combinations in 35 different ways (exhaustively). Then, we randomly pick one possible workload for each category and
carry out simulations for all the 35 different workloads. We use the normalized weighted IPC ~\cite{eyerman2013restating}.of the eight cores as the performance metric. The formula for this metric is as follows. We measure the IPC of each individual benchmark on an isolated machine. Then, we divide the IPC of benchmark of a multiprogramming mix by its corresponding individual IPC (weighting) and then take the average of the resultant normmalized IPCs.

\begin{figure}[t]
\begin{subfigure}{\linewidth}
  \includegraphics[trim=0 0 0 40, clip, width=\linewidth]{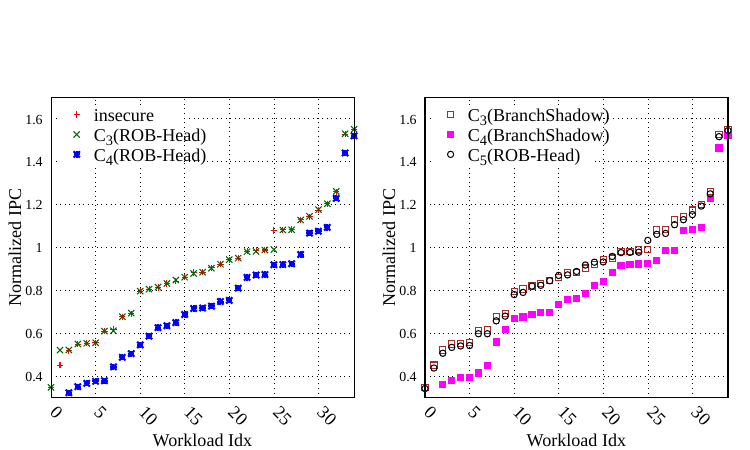}
  \\[-3ex]
  \caption{ \textmd{The performance of results for multiprogramming configurations, using the normalized weighted IPC metric.}}
  \label{fig:spec_multicore_ipc}
\end{subfigure}
\\[-5ex]
\begin{subfigure}{\linewidth}
  \includegraphics[trim=0 88 0 0, clip, width=\linewidth]{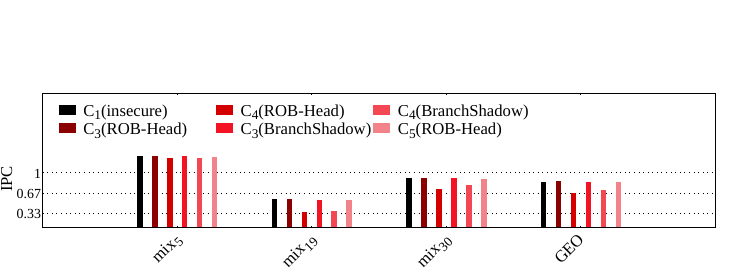}
  \\[-3ex]
  \caption{ \textmd{The IPC comparison in three particular workload mixes, namely, $mix_5$, $mix_{19}$ and $mix_{30}$. The geometric mean IPC over all workload mixes is also shown.}}
  \label{fig:spec_ipc_selected}
\end{subfigure}
\\[-6ex]
\begin{subfigure}{\linewidth}
  \includegraphics[trim=0 88 0 0, clip, width=\linewidth]{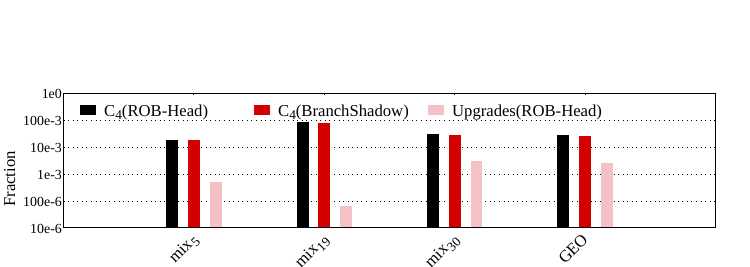}
  \\[-4ex]
  \caption{ \textmd{The DSRM fraction comparison for three workload mixes, namely, $mix_5$, $mix_{19}$ and $mix_{30}$. We also show the fraction of demand accesses that cause upgrades from S to M state.}}
  \label{fig:spec_redos}
\end{subfigure}
\caption{\textmd{SPECrate 2017 multiprogramming results (see \S\ref{subsection:spec_multiprogramming}).} }
\end{figure}

\textbf{Performance Results.}
Figure~\ref{fig:spec_multicore_ipc} shows the performance results of the multiprogramming simulations for all 35 mixes. The simulation results for each of the five configurations are sorted in increasing order of normalized weighted IPC.  Figure~\ref{fig:spec_ipc_selected} shows IPC in three particular cases, one where the IPC degradation due to DSRM is high and one where it is relatively low and one where it is moderate. Figure~\ref{fig:spec_redos} shows the fraction of cache demand accesses that result in DSRM redos. This is helpful for understanding the performance of the multiprogrammed system. It also shows the fraction of demand accesses which have S to M upgrades.

\emph{Average Results.}  
The performance results for $C_1$(insecure), $C_3$(ROB-Head), $C_4$(ROB-Head), $C_3$(BranchShadow), $C_4$(BranchShadow) and $C_5$ (ROB-Head) are 0.85, 0.85, 0.66, 0.85, 0.71 and 0.83. The average DSRM fraction for $C_4$(ROB-Head), $C_4$(BranchShadow) and the average upgrade fraction for $C_5$(ROB-Head) are 0.027, 0.025 and 0.0025.
Thus, we observe an average performance overhead of  
32\% for the ROB-Head model and 26\% for the BranchShadow model. The average performance overheads are 2\% for T + DC + SS-MESI variant under the ROB-Head model. 
 
\subsubsection{Examining Particular Mixes} 
In $mix_{5}$, the benchmarks that run together are \texttt{bwaves}, \texttt{lbm}, \texttt{lbm}, \texttt{bwaves}, \texttt{exchange}, \texttt{perlbench}, \texttt{imagick} and \texttt{x264}. Most of these benchmarks (other than \texttt{x264})  are on the lower side of DSRM redo fractions. The performance results for $C_1$(insecure), $C_3$(ROB-Head), $C_4$(ROB-Head), $C_3$(BranchShadow), $C_4$(BranchShadow) and $C_5$(ROB-Head) are 1.26, 1.26, 1.23, 1.26, 1.23 and 1.25. The performance degradation is under 3\% for this group.

In $mix_{19}$, the benchmarks that run together are \texttt{fotonik}, \texttt{fotonik}, \texttt{parest}, \texttt{parest}, \texttt{roms}, \texttt{roms}, \texttt{bwaves} and \texttt{lbm}. These benchmarks have overall a very high amount of DSRM redos. Compared to the average  DSRM redo fraction (2.6\%), the average of this group is 8\%, which is correlated with the overall lower performance result. 
The performance results for $C_1$(insecure), $C_3$(ROB-Head), $C_4$(ROB-Head), $C_3$(BranchShadow), $C_4$(BranchShadow) and $C_5$ (ROB-Head) are 0.55, 0.55, 0.35, 0.55, 0.36 and 0.54. Thus, the  performance overhead for this group is under 37\%.

In $mix_{30}$, the benchmarks that run together are \texttt{cam4}, \texttt{mcf}, \texttt{parest}, \texttt{cactus}, \texttt{omnetpp}, \texttt{omnetpp}, \texttt{bwaves}, \texttt{bwaves} and \texttt{x264}. Out of these,  \texttt{mcf}, \texttt{parest} and \texttt{omnetpp} have a  significantly higher DSRM redo fraction compared to other benchmarks but the others are not on the higher side in terms of DSRM redo counts. The average DSRM redo count of this group is 2.8\%, which is close to the average (2.7\%). Hence, we do expect a higher performance deterioration compared to other benchmarks. The performance results for $C_1$(insecure), $C_3$(ROB-Head), $C_4$(ROB-Head), $C_3$(BranchShadow), $C_4$(BranchShadow) and $C_5$(ROB-Head) are 0.90, 0.90, 0.73, 0.90, 0.78 and 0.89. The performance overhead of this group is under 22\%. This is significantly better than $mix_{19}$ and significantly worse than $mix_{5}$, which is what we would expect based on the DSRM redo count trends.

\captionsetup[subfigure]{%
   justification=raggedright,
   labelfont=bf,
  singlelinecheck=off
}

\section{Discussion and Related Work}
\label{section:discussion}
 
Firstly, we consider other coherence protocols that have an E state (see \S\ref{subsection:other_E_protocols}). Secondly, we make the case that other defenses need to be carefully inspected for microarchitectural defense assumption violations in a holistically secure microarchitecture (see \S\ref{subsection:impact}).

\subsection {Security Issues With Related Coherence Protocols}
\label{subsection:other_E_protocols}
\textcolor{black} { The microarchitectural defense assumption violation in this work that permits  coherence information to flow from the cache to the core 
also applies in the context of any coherence protocol that can create a remote E state cache line. For example, DSRC creates a selective redo on the E state, which can be used to detect remote cache lines. 
Common examples of other protocols with the remote E state are MOESI and MESIF~\cite{sorin2011primer}. They are  optimized versions of the baseline MESI protocol. A more security-oriented version of the MESI protocol makes E state and S state accesses take an equal amount of time.
However, DSRC delays on remote E coherence state still apply~\cite{yao2019covert}.
Some defense schemes~\cite{miao2022swiftdir} rely on OS provided shared read-only memory information on the page table/TLB to avoid E state for shared data or code. It does not cover other shared read-only memory, for example, shared read-only objects used by two Javascript sandboxes~\cite{isolatedVM}. }

\textcolor{black}{ 
\textbf{Difference between SS-MESI and MSI.} The SS-MESI protocol permits dirty cache lines to return to an E state eventually, if the underlying MESI protocol supports it (e.g., GEM5's underlying MESI protocol). On the other hand, MSI, does not permit a return to E state in any situation, so it is less efficient with respect to writeback. Overall, the protocols operate similarly other than this difference. Hence, we favor SS-MESI by a small advantage compared to MSI.}

\subsection{Implications For Other Defenses}
\label{subsection:impact}

\emph{Defenses That Use TORC.} \textcolor{black}{Secure non-inclusive caches~\cite{yan2019secdir} use a TORC variant to protect remote E state. When a remote cache line is detected, data is fetched from memory instead of getting it from the remote cache.} 
TimeCache~\cite{ojha2021timecache} attaches process-context to the cache lines, which helps it to reduce the number of timing delays. However, in both variants, E states can still  cause redos, so DSRC continues to enable shield bash attacks. 

\emph{Defenses That Use DSRC. } Many speculative defenses~\cite{ainsworth2021ghostminion,ainsworth2020muontrap,saileshwar2019cleanupspec} use DSRC along with speculative buffering. The speculative buffer does not prevent the creation of remote E states.  Some techniques use randomization~\cite{saileshwar2019cleanupspec} along with DSRC but this also does not offer any protection of the remote E state. Hence, these combinations are also vulnerable to shield bash. 

\textbf{Other Defenses. } There are a host of other microarchitectural defenses~\cite{kiriansky2018dawg,saileshwarbespoke,dessouky2021chunked,giner2022scatter,qureshi2018ceaser,saileshwar2021mirage,unterluggauer2022chameleon,dessouky2020hybcache,ainsworth2021ghostminion,ainsworth2020muontrap,yan2018invisispec,bahmani2021cure,lee2019keystone} for other microarchitectural attacks than those discussed so far~\cite{purnalprime+,briongos2020reload+,wan2021volcano,chen2021leaking,pessl2016drama,khaliq2021timing,tan2021invisible,gruss2016rowhammer,mutlu2019rowhammer,murdock2020plundervolt,skarlatos2021jamais}. Attacks are mitigated by defensively modifying both cache~\cite{omar2020ironhide,saileshwarbespoke,dessouky2021chunked,ramkrishnan2024non,giner2022scatter} and non-cache~\cite{loughlin2021dolma,harris2019cyclone,anagnostopoulos2018overview,khaliq2021timing,murdock2020plundervolt} components. 
We believe that it is an important  area of future research to study defenses (and any accompanying defense assumption violations) for a holistic and secure microarchitecture design. This will help to discover and mitigate shield bash-like attacks in more general scenarios.

\textcolor{black}{
\subsection{Future Work: Other Possible MDAVs}
\label{section:future_work}
}
\textcolor{black}{
\textbf{Declassification and Timing Obfuscation.} 
One of the important defensive optimizations  is the declassification optimization for secure speculation  schemes~\cite{aimoniotis2023recon,choudhary2021speculative}. Declassification adds metadata to the cache lines which also potentially travels between cores~\cite{aimoniotis2023recon}, affecting the pipeline state. This could potentially interfere with timing obfuscations in a similar manner as the coherence related MDAV. }

\textcolor{black}{
\textbf{Declassification, Randomization and Isolation.} Declassification has a potential MDAV with respect to both randomization~\cite{unterluggauer2022chameleon,song2021randomized} and partitioning~\cite{yan2019secdir,ramkrishnan2024non} defenses, which are cache-based defenses. Declassification causes certain cache data to be declared as non-protected. Speculative side-effects that reveal this data are allowed by the processor.  
This can be abused by an attacker because it can create speculative side-effect based covert channels. For example, a sandboxed attacker in one thread may abuse declassification to create speculative covert channels to another sandbox in a different thread, based on modifying state that is visible across partitions or randomizations, e.g., coherence states. }
\textcolor{black}{
One possible mitigation is to also add in  protections (such as DSRC) against potential speculative transmitters of declassified data. TORC can also be used in conjunction with partitioning~\cite{yan2019secdir} or randomization~\cite{ramkrishnan2020first}, so the interactions TORC and DSRC are important considerations.}

\textcolor{black} {
\subsection{Future Work: Automating MDAV Detection}
For automation, one possibility is to create rules that need to be satisfied for different defenses to work correctly. This is often possible using existing formal model-checkers like Alloy~\cite{vakili2012temporal}.  Unmodified TORC has a rule that it always does a delay on remote cache hits  whereas DSRC has a rule that it always does redos on remote speculative hits.  
If non-interference invariants are also defined and applied to the TORC + DSRC system, 
it is plausible that it could trigger the counterexample discussed in this paper. We leave the creation of a general automation framework to future work.}

\section{Conclusion}
\label{section:conclusion}
We examine an interaction between two  classes of state-of-the-art microarchitectural defenses, \emph{timing obfuscation of remote cache lines} (TORC) and \emph{delaying 
 speculative changes to remote cache lines} (DSRC). In the integrated defense, DSRC can be abused to create selective secret-dependent redo operations in the core by leveraging coherence information, thus bypassing TORC protections. Attack simulations using GEM5 demonstrate the creation of a new cache-hit covert channel using remote E cache lines.
We  propose two possible fixes, \emph{delay speculative on remote and miss} (DSRM) and \emph{Start-With-S MESI} (SS-MESI).
DSRM uses additional \emph{redo} accesses on cache misses to equalize the timings between cache misses and remote cache line hits, whereas, SS-MESI inserts S state on load misses in the LLC, thus eliminating redos.
Performance measurements on SPECrate 2017 benchmaks and PARSEC benchmarks show that the repair costs are less than 2.8\% for TORC + DSRC + SS-MESI and less than 32\% for TORC + DSRM.

\bibliographystyle{ACM-Reference-Format}
\bibliography{sample-base}

\end{document}